\documentclass[10pt,conference]{IEEEtran}
\IEEEoverridecommandlockouts
\usepackage{cite}
\usepackage{amsmath,amssymb,amsfonts}
\usepackage{algorithmic}
\usepackage{graphicx}
\usepackage{textcomp}
\usepackage{xcolor}
\usepackage[a4paper, total={184mm,239mm}]{geometry}
\def\BibTeX{{\rm B\kern-.05em{\sc i\kern-.025em b}\kern-.08em
    T\kern-.1667em\lower.7ex\hbox{E}\kern-.125emX}}

\newtheorem{definition}{Definition}
\newtheorem{theorem}{Theorem}
\renewcommand{\IEEEQED}{\IEEEQEDopen}
\usepackage{mathtools}
\usepackage{tikz}
\usetikzlibrary{shapes.geometric,arrows,automata}
\usepackage{varwidth}
\usepackage{multirow}
\usepackage{subcaption}
\usepackage{xurl}

\begin{document}

\title{
	Active Learning of Abstract System Models 
	\\from Traces
	using Model Checking [Extended]\\
\thanks{This research was funded in part by the Semiconductor Research Corporation,~Task
	2707.001, and Balliol College, Jason Hu~scholarship.}
}

\author{\IEEEauthorblockN{1\textsuperscript{st} Natasha Yogananda Jeppu}
\IEEEauthorblockA{\textit{University of Oxford} \\
Oxford, UK\\
natasha.yogananada.jeppu@cs.ox.ac.uk}
\and
\IEEEauthorblockN{2\textsuperscript{nd} Tom Melham}
\IEEEauthorblockA{\textit{University of Oxford} \\
Oxford, UK\\
tom.melham@cs.ox.ac.uk}
\and
\IEEEauthorblockN{3\textsuperscript{rd} Daniel Kroening}
\IEEEauthorblockA{\textit{Amazon, Inc} \\
 London, UK\\
daniel.kroening@magd.ox.ac.uk}
}


\maketitle

\begin{abstract}
We present a new active model-learning approach to generating abstractions of a
system implementation, as finite state automata (FSAs), from execution traces.
Given an implementation and a set of observable system variables, the generated
automata admit all system behaviours over the given variables and provide useful
insight in the form of invariants that hold on the implementation. To achieve
this, the proposed approach uses a pluggable model learning component that can
generate an FSA from a given set of traces. Conditions that encode a
completeness hypothesis are then extracted from the FSA under construction and
used to evaluate its degree of completeness by checking their truth value
against the system using software model checking. This generates new traces that
express any missing behaviours. The new trace data is used to iteratively refine
the abstraction, until all system behaviours are admitted by the learned
abstraction. To evaluate the approach, we reverse-engineer a set of publicly
available Simulink Stateflow models from their C implementations.
\end{abstract}


\section{Introduction}
Execution traces provide an exact representation of the system behaviours that
are exercised when an instrumented implementation runs. This is leveraged by
automated model learning algorithms to generate system abstractions.

Modern passive learning algorithms also infer guards and operations on system
variables from the trace data, yielding symbolic models~\cite{model_daikon,
	compute_walkinshaw, Walkinshaw2016, jeppu}. But the behaviours admitted by these
models are, of course, limited to only those manifest in the traces. So
capturing all system behaviour by the generated system models is conditional on
devising a software load that exercises all relevant system behaviours.

This can be difficult to achieve in practice, especially when a system comprises
multiple components and it is not obvious how each component will behave in
conjunction with the others. Random input sampling is a pragmatic choice in this
scenario, but it does not guarantee that generated models admit all system
behaviour. This is discussed further in Section~\ref{sec:random}.

On the other hand, active learning algorithms can, in principle, generate exact
system models~\cite{lstar,mat_star}. But when used~in practice to learn symbolic
abstractions, these algorithms suffer from high query complexity and can only
learn models with transitions labelled by simple predicates, such as
equality/inequality relations~\cite{grey_box_sl, Howar2018ActiveAL, Howar2019}. 

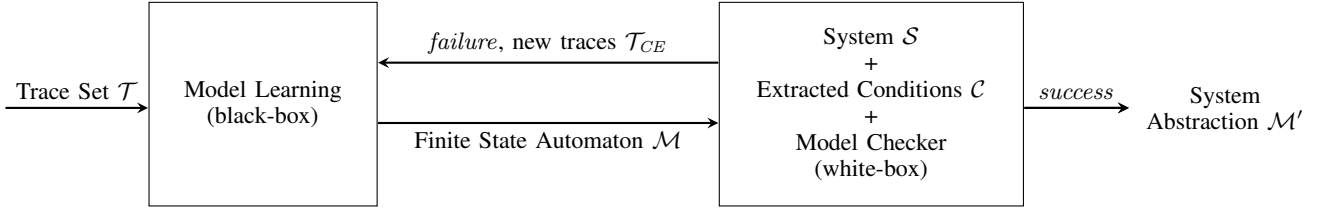
\begin{figure*}[t]
	\small
	\centering
	\begin{tikzpicture}act
	\draw (-1,1.2) rectangle (2,3.9) node[pos=.5,align=center] {Model Learning \\(black-box)};
	\draw (6.5,1.2) rectangle (10.5,3.9) node[pos=.5,align=center] {System $\mathcal{S}$ \\+ \\Extracted Conditions $\mathcal{C}$ \\+ \\ Model Checker \\ (white-box)};
	\node (A) at (-3,2.5) {};
	\node (B) at (-0.9,2.5) {};
	\node (C) at (1.9,2.3) {};
	\node (D) at (6.6,2.3) {};
	\node (I) at (6.6, 3.1) {};
	\node (J) at (1.9, 3.1) {};
	\node (O) at (10.4, 2.5) {};
	\node (P) at (12, 2.5) {};
	\draw [thick, ->, >=stealth] (A) to node[anchor=south] {Trace Set $\mathcal{T}$} (B);
	\draw [thick, ->, >=stealth] (C) to node[anchor=north] {Finite State Automaton $\mathcal{M}$} (D);
	\draw [thick, ->, >=stealth] (I) to node[anchor=south] {$\mathit{failure}$, new traces $\mathcal{T}_{\mathit{CE}}$} (J);
	\draw [thick, ->, >=stealth] (O) to node[anchor=south] {$\mathit{success}$} (P);
	\draw [thick, ->, >=stealth] (O) to node[anchor=west,align=center,xshift = 8mm] {System \\ Abstraction $\mathcal{M}'$} (P);
	\end{tikzpicture}
	\caption{Overview of the active model-learning algorithm.}
	\label{fig:active}
\end{figure*}

We present a new active learning approach to derive system abstractions, as
finite state automata (FSA), of a system implementation instrumented to observe
a set of system variables.  The generated abstractions admit all system
behaviours over these variables and provide useful insight in the form of
invariants that hold on the implementation. As illustrated in
Fig.~\ref{fig:active}, the approach is a grey-box algorithm. It combines a
black-box analysis, in the form of model learning from traces, with a white-box
analysis, in the form of software model checking~\cite{mc-book}. The model
learning component can be any~algorithm that can generate an FSA that accepts a
given set of system execution traces. Model checking is used to evaluate the
degree of completeness of the learned automaton and identify any missing
behaviours.

The core of this new approach is as follows. Given an instrumented system
implementation, a set of observable variables and a model learning algorithm,
the structure of a candidate abstraction generated by model learning is used to
extract conditions that collectively encode the following completeness
hypothesis: For any transition in the system defined over the set of
observables, there is a corresponding transition in the generated abstraction.
To verify the hypothesis, all conditions are checked against the implementation
using model checking. Any violation indicates missing behaviours in the
candidate model. This evaluation procedure operates at the level of the
abstraction and not individual system traces, unlike query-based active learning
algorithms, and therefore can be easily implemented using existing model
checkers.

The procedure to evaluate degree of completeness of the learned model yields a
set of new traces that exemplify system behaviours identified to be missing from
the model. New traces are used to augment the input trace set for model learning
and iteratively refine the abstraction until all conditions are satisfied. The
resulting learned model is the system abstraction that has been proven to admit
all system behaviours defined over the set of observable variables. Further, the
conditions extracted from the final generated abstraction serve as invariants
that hold on the implementation. Given a model learning algorithm that can infer
symbolic abstractions from trace data, such as~\cite{jeppu}, the approach can
learn models that are more expressive than the abstractions learned using
existing active learning algorithms.

\section{Background}
\subsection{Formal Model}
The system for which we wish to generate an abstraction is represented as a
tuple $\mathcal{S} = (X, X', R, \mathit{Init})$.  $X = \{x_{1}, \dots, x_{m}\}$
is a set of observable system variables over some domain~$D$ that can be used to
collect execution traces.  We simplify the presentation by assuming all
variables have the same domain.  The set $X' = \{x_{1}', \dots, x_{m}'\}$
contains corresponding primed variables, also over the domain~$D$. A primed
variable $x_i'$ represents an update to the unprimed variable~$x_i$ after a
discrete time step.  The transition relation $R(X, X')$ describes the
relationship between $x_i$ and $x_i'$ for $1 \leq i \leq m$ and is represented
using a characteristic function, i.e., a Boolean-valued expression over $(X \cup
X')$.  The set of initial system states is represented using its characteristic
function $\mathit{Init}(X)$.

A~\emph{valuation} $v: X \to D$ maps the variables in $X$ to values in~$D$. An
\textit{observation} at discrete time step $t$ is a valuation of the variables
at that time, and is denoted by~$v_{t}$.  A \textit{trace} is a sequence of
observations over time; we write a trace $\sigma$ with $n$ observations as a
sequence of valuations $\sigma = v_{1}, v_{2}, \dots, v_{n}$. We define an
execution trace or \emph{positive} trace for $\mathcal{S}$ as a trace $\sigma =
v_{1}, \dots, v_{n}$ that corresponds to a system execution path, i.e.,
$(v_t,v_{t+1}) \models R$ for $1 \leq t < n$ and there exists a valuation $v'
\models \mathit{Init}$ such that $(v',v_1) \models R$. A \emph{negative} trace
is a trace that does not correspond to any system execution path. We represent
the set of execution traces by $\mathit{Traces}_X(\mathcal{S})$.

\begin{figure}[b]
	\centering
	\small
	\begin{tikzpicture}[->,>=stealth',shorten >=1pt,auto,node distance=1.69cm,thick]
	\tikzstyle{every state}=[fill=white,draw=black,text=black]
	\node[initial,state]         (B) {$\mathbf{q_1}$};
	\node[state]         (C) [below of=B] {$\mathbf{q_2}$};
	\path 
	(B) edge [loop right]             node [anchor=west] {$(s'{=}\textrm{ Off})$} (B)
	(B) edge [bend left]             node [anchor=west,align=center] {$ (\mathit{inp.temp} > \textrm{T\_thresh})$\\${\wedge}(s'{=}\textrm{ On})$} (C)
	(C) edge [loop right]             node [anchor=west] {$(s'{=}\textrm{ On})$} (C)
	(C) edge [bend left]             node [anchor=east,align=center] {$\neg (\mathit{inp.temp} > \textrm{T\_thresh})$\\${\wedge}(s'{=}\textrm{ Off})$} (B);
	
	\end{tikzpicture}
	\caption{Generated abstraction modeling operation mode switches for a Home Climate-Control Cooler system using~\cite{t2m}.}
	\label{fig:example_automata}
\end{figure}
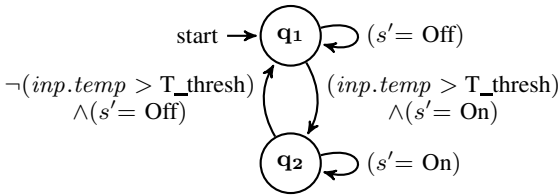 

The active learning algorithm learns a system model as an FSA. Our abstractions
are represented symbolically and feature predicates on the transition edges,
such as the abstraction in Fig.~\ref{fig:example_automata}, and therefore extend
finite automata to operate over infinite alphabets. We represent the learned
abstraction as a non-deterministic finite automaton~(NFA)
$\mathcal{M}\allowbreak=(\allowbreak\mathcal{Q},\allowbreak
\mathcal{Q}^0,\allowbreak\Sigma, \allowbreak{F}, \allowbreak\delta)$ over an
infinite alphabet $\Sigma$, where $\mathcal{Q}$ is a finite set of states,
$\mathcal{Q}^0 \subseteq \mathcal{Q}$ are the initial states,
${F}\subseteq\mathcal{Q}$ is the set of accepting states, and $\delta:
\mathcal{Q} \times \Sigma \rightarrow \mathcal{P(Q)}$ is the transition
function. The alphabet $\Sigma$ corresponds to the set of valuations for
variables in~$X$, i.e., $\Sigma = (X \longrightarrow D)$. The exact alphabet
need not be known a priori.

The NFA admits a trace $\sigma = v_{1}, \dots, v_{n}$ if there exists a sequence
of automaton states $q_{1}, \dots, q_{n{+}1}$ such that $q_{1} \in
\mathcal{Q}^0$ and $q_{i+1} \in \delta(q_{i},v_{i})$ for $1 \leq i \leq n$. Any
finite prefix of a system execution trace $\sigma$ is also~an execution trace.
Thus, if the generated NFA admits $\sigma$, it must also admit all finite
prefixes of $\sigma$. In other words, the language of the automaton,
$L(\mathcal{M})$, must be \textit{prefix-closed}. 

\begin{definition}[Prefix-Closure] \label{nfa_definition} %
	A language $L$ is said to be prefix-closed if for all words $w \in L$,
$\mathit{Pref}(w) \subseteq L$. Here, $\mathit{Pref}(w)$ denotes the set of all
prefixes of the word $w$.
\end{definition}

All states of our automaton are accepting, i.e., the NFA rejects traces by
running into a `dead end'.
\subsection{Model Learning from Execution Traces}\label{sec:t2m}

The approach uses a pluggable model learning component~to
generate models from traces.  Our requirement for this component is simple:
given a set of execution traces $\mathcal{T}$, the component returns an NFA
$\mathcal{M}$ that accepts (at~least) all traces in $\mathcal{T}$.

There are several model-learning algorithms that satisfy this
requirement~\cite{model_daikon, efsm_state_merge, compute_walkinshaw,
	Walkinshaw2016, jeppu, Biermann:1972:SFM:1638603.1638997}. 
For our experiments, we use the
automata learning algorithm in~\cite{jeppu} as the model learning component. The
algorithm uses a combination of Boolean Satisfiability (SAT) and program
synthesis~\cite{gulwani2017program} to generate compact accurate symbolic
models using only
execution traces. The algorithm has been implemented as an open source tool,
Trace2Model (T2M)~\cite{t2m}, which we use for our experiments.

Our choice of algorithm allows us to demonstrate the effectiveness of the
active learning algorithm with minimal assumptions. 
T2M uses \emph{only} execution traces to generate models, unlike other
algorithms that use a priori known LTL properties~\cite{model_SAT, exact_fsm,
	state_merge}. Furthermore, T2M can generalise over multiple variables to
generate symbolic models with predicates on the transition edges
(Fig.~\ref{fig:example_automata}). By demonstrating that the active learning
algorithm can be used with T2M, we show that it can also be applied to learn
models with transition edges labelled with simple Boolean events or letters from
the alphabet $\Sigma$.


\section{Active Learning of Abstract System Models}

The NFA $\mathcal{M}$ generated by the model learning component admits all
system behaviours captured by the input trace set $\mathcal{T}$.  However,
$\mathcal{T}$ might not capture all system behaviour. To evaluate the degree of
completeness of the set of traces, we use the structure of the NFA $\mathcal{M}$
to extract conditions~that can be checked against the system implementation. 
The conditions collectively encode the following completeness hypothesis: For
any transition available in the system defined by the transition relation $R$,
there is a corresponding transition in $\mathcal{M}$.

The hypothesis is formulated based on defining a \textit{simulation} relation
between the system $\mathcal{S}$ and abstraction $\mathcal{M}$.
\begin{definition} If $\mathcal{Q}'$ represents the set of system states for
	$\mathcal{S}$ and $q'_{v_t} \in \mathcal{Q}'$ represent the system state
	characterised by valuation $v_t$ of variables in $X$, then we define a binary
	relation $\mathcal{R}' \subseteq \mathcal{Q}' \times \mathcal{Q}$ to be a
	simulation if $(q'_{v_t},q_i) \in \mathcal{R}'$ implies that $\forall
	(v_t,v_{t+1}) \models R$ i.e, $q'_{v_t} \rightarrow q'_{v_{t+1}}$, $\exists
	q_{i+1} \in \mathcal{Q}$ such that $q_{i+1} \in \delta(q_i, v_{t+1})$ and
	$(q'_{v_{t+1}},q_{i+1}) \in \mathcal{R}'$.
\end{definition}

If such a relation $\mathcal{R}'$ exists, we denote $\mathcal{S} \preceq \mathcal{M}$. A
consequence of the existence of a simulation relation is trace inclusion, i.e.,
$\mathcal{S} \preceq \mathcal{M} \implies \mathit{Traces}_X(\mathcal{S})
\subseteq L(\mathcal{M})$, as will be proved later in this section.

The extracted conditions are checked against the system implementation using
software model checking (See Section~\ref{sec:k-ind}.). The procedure returns
\textit{success} if all conditions are satisfied and \textit{failure}, along
with a counterexample trace, if there is a violation. The counterexample trace
is a sequence of valuations of variables in $X$ that corresponds to an execution
path in the system representing the missing behaviour. The set of counterexample
traces $\mathcal{T}_{\mathit{CE}}$ obtained for all violations is used to
augment the input trace set i.e., $\mathcal{T} \gets \mathcal{T} \cup
\mathcal{T}_{\mathit{CE}}$. This is then fed to the model learning component to
generate an extended abstraction that covers missing behaviours.

When all conditions are satisfied, the algorithm returns the generated automaton
as the final learned system abstraction $\mathcal{M}'$. The conditions extracted
from $\mathcal{M}'$ serve as invariants that hold on the system implementation.



\subsection{Completeness Conditions for a Candidate Abstraction}

Given a candidate abstraction $\mathcal{M}$ for a system $\mathcal{S}$, we
extract the following conditions:
%
\begin{equation}
v_t \models \mathit{Init}\wedge (v_t,v_{t+1}) \models R \implies \bigvee_{p_o \in
	P_{(0,out)}} v_{t+1} \models p_o\label{eq:prop1}\tag{1}
\end{equation}
%
where $P_{(0,out)}$ is the set of predicates for all outgoing transitions from
an automaton state $q_0 \in \mathcal{Q}^0$, and for all $p_i \in P_{(j,in)}$
%
\begin{equation}
v_t  \models p_i\wedge (v_t,v_{t+1}) \models R
\implies \bigvee_{p_o \in P_{(j,out)}} v_{t+1} \models p_o
\label{eq:prop2}\tag{2}
\end{equation}
%
where $P_{(j,in)}$ is the set of predicates on the incoming transitions to state
$q_j \in \mathcal{Q}$ and $P_{(j,out)}$ is the set of predicates on outgoing
transitions from $q_j$. A~condition of the form~\eqref{eq:prop2} is extracted
for every state $q_j \in \mathcal{Q}$.

We compute the fraction of conditions that hold on the system, denoted by
$\alpha$, as a quantitative measure of the degree of completeness of the
learned model.  If all extracted conditions hold, i.e., $\alpha = 1$, then
the generated model admits all system behaviours.  A~violation indicates
missing behaviour in $\mathcal{M}$.

\begin{theorem}\label{thm:1}
	Given a candidate abstraction
$\mathcal{M}\allowbreak=(\allowbreak\mathcal{Q},\allowbreak
\mathcal{Q}^0,\allowbreak\Sigma, \allowbreak{F}, \allowbreak\delta)$ for a
system $\mathcal{S} = (X, X', R, \mathit{Init})$ and the set of conditions
$\mathcal{C}$ extracted from $\mathcal{M}$,
	\begin{align*}
	\bigwedge_{r \in \mathcal{C}} r \implies
	\mathit{Traces}_{X}(\mathcal{S}) \subseteq L(\mathcal{M})
	\end{align*}
\end{theorem}

\begin{IEEEproof}
We will prove this by contradiction. Let us assume that all conditions are
satisfied and there exists a trace $\sigma = {v_1, v_2, \dots, v_n} \in
\mathit{Traces}_{X}(\mathcal{S})$ such that $\sigma \not\in L(\mathcal{M})$. Let
$\sigma' = v_1, v_2, \dots, v_{j}$ be the longest prefix of $\sigma$ that is
accepted by $\mathcal{M}$. Hence, there exists a sequence of states $q_1, q_2,
\dots, q_{j{+}1}$ such that $q_1 \in \mathcal{Q}^0$ and $q_{i+1} \in
\delta(q_i,v_i)$, for $1 \leq i \leq j$.

If $j > 0$, there exists a predicate $p_i \in P_{(j+1,in)}$ such that $v_{j}
\models p_i$, $(v_j,v_{j+1}) \models R$ as $\sigma \in
\mathit{Traces}_{X}(\mathcal{S})$, and $v_{j{+}1} \not\models p_o, \allowbreak
\forall p_o \in P_{(j+1,out)}$. This violates condition~\eqref{eq:prop2} and
therefore, contradicts our assumption. If $j = 0$, since $\sigma \in
\mathit{Traces}_X(\mathcal{S})$, there exists a valuation $v' \models
\mathit{Init}$ such that $(v',v_1) \models R$, and $v_1 \not\models p_o,
\allowbreak \forall p_o \in P_{(0,out)}$. This is a violation of
condition~\eqref{eq:prop1} and therefore, contradicts our assumption.
\end{IEEEproof}

%
%
 
\subsection{Verifying Extracted Conditions Against the System}\label{sec:k-ind}

To enable the application of existing software model checkers, we construct
source code for functions that encode conditions~\eqref{eq:prop1}
and~\eqref{eq:prop2} of the form $v_t \models r \wedge (v_t,v_{t+1})\models R
{\implies} v_{t+1} \allowbreak {\models} \, s$ as assume/assert pairs, as
illustrated in Fig.~\ref{fig:verify}. Here, $X' = f(X)$
(line~\ref{line_transition} in Fig~\ref{fig:verify}) represents an
implementation of the transition relation $R$. System behaviours are modelled as
multiple unwindings of the loop in line~\ref{start_loop} in
Fig.~\ref{fig:verify}.

To check if the system satisfies a condition, we run model checking using
$k$-induction~\cite{k-ind,dhkr2011-sas} on the function with $k = 1$, as each
condition~describes a single system transition. Note that the procedure in
Fig.~\ref{fig:verify} does not start from an initial state, but an arbitrary
state that satisfies~$r$.  A~proof with $k=1$ is thus sufficient to assert that
the system satisfies the condition for any number of transitions. When all
assume/assert pairs are proved valid, this implies that the extracted conditions
are always satisfied and therefore can be used as invariants that hold on the
system. Other model checking algorithms can be used in place of
$k$-induction for the condition check.

\begin{figure}[b]
	\centering
	\small
	\begin{minipage}{0.45\columnwidth}
		\centering
		\begin{algorithmic}[1]
			\STATE \textbf{assume}($r$)
			\WHILE {true}\label{start_loop}
			\STATE $X' = f(X)$\label{line_transition}
			\ENDWHILE
			\STATE \textbf{assert}($s$)
		\end{algorithmic}
		\subcaption{}
		\label{fig:verify}
	\end{minipage}
	\quad
	\begin{minipage}{0.45\columnwidth}
		\centering
		\begin{algorithmic}[1]
			\STATE \textbf{assume}($\mathit{Init}(X)$)
			\WHILE{true}
			\STATE $X' = f(X)$
			\ENDWHILE
			\STATE \textbf{assert}($\neg s'$)
		\end{algorithmic}
		\subcaption{}
		\label{fig:ce_check}
	\end{minipage}
	\caption{(a) Condition check (b) Spurious counterexample check.}
	\label{fig:wrapper}
\end{figure}

In case of a failure, the checker returns a sequence of valuations $\sigma'' =
v_t, v_{t+1}$ as the counterexample, such that $v_t \models r \wedge
(v_t,v_{t{+}1}) \models R \wedge v_{t{+}1} \not\models s$. This can be
used to construct a set of new traces as follows. For each trace $\sigma \in
\mathcal{T}$ we find the smallest prefix $\sigma' = v_1, v_2, \dots, v_j$ such
that $v_j \models r$. We then construct a new trace $\sigma_{\mathit{CE}} = v_1,
v_2, \dots, v_{j{-}1}, v_t, v_{t{+}1}$ for each prefix $\sigma'$. Note that
since $v_t \models r$, the new trace $\sigma_{\mathit{CE}}$ does not change the
system behaviour represented by $\sigma'$ but merely augments it to include the
missing behaviour. The set of new traces $\mathcal{T}_{\mathit{CE}}$ thus
generated is used as an additional input to the model learning component, which
in turn generates a refined abstraction that admits the missing behaviour.

For a violation of condition~\eqref{eq:prop1}, the checker returns a
counterexample $\sigma'' = v_0, v_1$ such that $v_0 \models \mathit{Init}$ and
$(v_0,v_1) \models R$. $\sigma''$ is therefore a valid counterexample. However,
the counterexample for a violation of condition~\eqref{eq:prop2} could be
spurious. Let $\sigma'' = v_t, v_{t+1}$ be the corresponding counterexample
generated by the model checker. Here, it is not guaranteed that the system state
characterised by $v_{t}$ is reachable from an initial system state. Therefore,
the counterexample may not actually correspond to missing system behaviour.

\subsection{Identifying Spurious Violations}
To check if a counterexample $\sigma'' = v_t, v_{t+1}$ is spurious, the valuation $v_t$ is
encoded as the following Boolean formula:
%
\begin{equation*}
s' \coloneqq \bigwedge_{x_i \in X} (x_i = v_{t}(x_i)) 
\end{equation*}
%
and the negation, $\neg s'$, is used to assert that $s'$ never holds at any
point in the execution of $\mathcal{S}$ starting from an initial state, as shown
in Fig.~\ref{fig:ce_check}.  We verify this using $k$-induction with $k > 1$. If
both the base case and step case for $k$-induction hold, it is guaranteed that
the counterexample is spurious, in which case we strengthen the assumption in
Fig.~\ref{fig:verify} to $(r \,\wedge \, \neg s')$ and repeat the condition
check.  In case of a violation only in the step case, there is no conclusive
evidence for the validity of the counterexample. Since we are not interested in
generating an exact model of the system but rather an over-approximation that
provides useful insight into the system, we treat such a counterexample as valid
but record it for future reference.

For the bound $k$, a value greater than or equal to the diameter of the system
guarantees completeness~\cite{k-ind}. In practice, it if often difficult to
determine this value without any system domain knowledge. An alternative is to
approximate the value of $k$ based on available trace information. For instance,
if there are observable counters in the system that affect system behaviour when
they reach a pre-defined limit, a good approximation for $k$ can be twice the
maximum counter limit. Note that a poor choice for the bound $k$ results in more
spurious behaviours being added to the model, resulting in low accuracy. But,
the learned models are guaranteed to admit all system traces defined over $X$,
irrespective of the value for~$k$.


\section{Evaluation and Results}\label{sec:exp}

\subsection{Evaluation Setup}
For our experiments we use T2M~\cite{t2m} as the model learning component, as
discussed in Section~\ref{sec:t2m}. To evaluate the degree of completeness we
use the C Bounded Model Checker (CBMC v5.35)~\cite{ckl2004}. We implement Python
modules for the following: generating the wrapper function to check each
condition, processing the CBMC~output to return the result of a condition check
and translating CBMC counterexamples into a set of trace inputs for model
learning. Note that any software model checker can be used in place of CBMC,
with relevant modules to process the corresponding outputs.

To evaluate the active learning algorithm, we attempt to reverse-engineer a set
of FSAs from their respective C implementations. For this purpose, we use the
dataset of Simulink Stateflow example models~\cite{stateflow}, available as part
of the Simulink documentation. The dataset comprises $51$ benchmarks that are
available in MATLAB 2018b. For each benchmark, we use Embedded
Coder~\cite{simcoder} to automatically generate a corresponding C code
implementation. The generated C implementation is used as the system
$\mathcal{S}$ in our experiments.

Out of the $51$ benchmarks, Embedded Coder fails to generate code for $7$; a
total of $13$ have no sequential behaviour and $3$ implement
Recursive~State~Machines (RSM)~\cite{rsm}.\footnote{We learn abstractions as
	FSAs, which are known to represent exactly the class of regular languages.
	Reverse-engineering an RSM from traces requires a modeling formalism that is
	more expressive than FSAs, such as Push-Down Automata (PDA)~\cite{pda}, which is
	outside the scope of this work.  In the future, we wish to look at extensions of
	this~work to generate RSMs.} We use the remaining $28$ benchmarks for our
evaluation.  The implementation and benchmarks are available online~\cite{exp}.

\begin{table*}[t]
	\centering
	\caption{Results of experimental evaluation of the active learning algorithm.}
	\label{tab:result}
	\begin{tabular}{|l|l|l|r|r|r|r|r|r|r|r|r|r|r|}
		\hline
		\multicolumn{3}{|c|}{\multirow{2}{*}{Benchmark}}                                                                                                                                                                           & \multicolumn{1}{c|}{\multirow{2}{*}{$|X|$}} & \multicolumn{1}{c|}{\multirow{2}{*}{$k$}} & \multicolumn{6}{c|}{Our Algorithm}                                                                                                                       & \multicolumn{3}{c|}{Random Sampling}                                                   \\ \cline{6-14} 
		\multicolumn{3}{|c|}{}                                                                                                                                                                                                     & \multicolumn{1}{c|}{}                       & \multicolumn{1}{c|}{}                     & $i$ & \multicolumn{1}{c|}{$d$} & \multicolumn{1}{c|}{$N$} & \multicolumn{1}{c|}{$\alpha$} & \multicolumn{1}{c|}{$T(s)$} & \multicolumn{1}{c|}{$\%T_{m}$} & \multicolumn{1}{c|}{$N$} & \multicolumn{1}{c|}{$\alpha$} & \multicolumn{1}{c|}{$T(s)$} \\ \hline
		\multicolumn{3}{|l|}{AutomaticTransmissionUsingDurationOperator}                                                                                                                                                           & 4                                           & 125                                       & 6   & \textbf{1}               & 5                        & \textbf{1}                    & 3678.3                      & 2.7                            & 4                        & 0.2                           & 38.2                        \\ \hline
		\multirow{2}{*}{\begin{tabular}[c]{@{}l@{}}BangBangControl\\ UsingTemporalLogic\end{tabular}}                          & \multicolumn{2}{l|}{Heater}                                                                       & \multirow{2}{*}{5}                          & \multirow{2}{*}{62}                       & 4   & \textbf{1}               & 4                        & \textbf{1}                    & 11845.5                     & 0.1                            & 3                        & 0.6                    & 64                          \\ \cline{2-3} \cline{6-14} 
		& \multicolumn{2}{l|}{On}                                                                           &                                             &                                           & 4   & \textbf{1}                      & 5                        & \textbf{1}                    & 11078                       & 0.2                            & 5                        & 0.7                           & 89.5                        \\ \hline
		\multicolumn{3}{|l|}{CountEvents}                                                                                                                                                                                          & 3                                           & 20                                        & 2   & \textbf{1}               & 3                        & \textbf{1}                    & 10.8                        & 41.7                           & 4                        & 0.8                   & 56.5                        \\ \hline
		\multicolumn{3}{|l|}{FrameSyncController}                                                                                                                                                                                  & 3                                           & 530                                       & 1   & 0                      & 1                        & 0                             & \multicolumn{2}{c|}{timeout}                                 & 2                        & 0.7                           & 31                          \\ \hline
		\multicolumn{3}{|l|}{HomeClimateControlUsingTheTruthtableBlock}                                                                                                                                                            & 7                                           & 10                                        & 1   & \textbf{1}               & 2                        & \textbf{1}                    & 5                           & 18.2                           & 2                        & \textbf{1}                    & 72.3                        \\ \hline
		\multirow{2}{*}{\begin{tabular}[c]{@{}l@{}}KarplusStrongAlgorithm\\ UsingStateflow\end{tabular}}                       & \multicolumn{2}{l|}{DelayLine}                                                                    & \multirow{2}{*}{5}                          & \multirow{2}{*}{100}                      & 2   & \textbf{1}               & 3                        & \textbf{1}                    & 430.9                       & 0.8                            & 3                        & \textbf{1}                    & 33.9                        \\ \cline{2-3} \cline{6-14} 
		& \multicolumn{2}{l|}{MovingAverage}                                                                &                                             &                                           & 3   & \textbf{1}               & 3                        & \textbf{1}                    & 1441.1                      & 0.4                            & 3                        & \textbf{1}                    & 35.2                        \\ \hline
		\multicolumn{3}{|l|}{LadderLogicScheduler}                                                                                                                                                                                 & 3                                           & 10                                        & 9   & \textbf{1}               & 4                        & \textbf{1}                    & 157                         & 63.9                           & 3                        & 0                          & 52.9                        \\ \hline
		\multicolumn{3}{|l|}{MealyVendingMachine}                                                                                                                                                                                  & 2                                           & 10                                        & 1   & \textbf{1}               & 4                        & \textbf{1}                    & 8.9                         & 49.1                           & 4                        & \textbf{1}                    & 67                          \\ \hline
		\multirow{5}{*}{\begin{tabular}[c]{@{}l@{}}ModelingACdPlayerradio\\ UsingEnumeratedDataType\footnotemark\end{tabular}} & \multirow{2}{*}{\begin{tabular}[c]{@{}l@{}}CdPlayer\\ BehaviourModel\end{tabular}} & DiscPresent  & \multirow{5}{*}{13}                         & \multirow{5}{*}{205}                      & 4   & 0.1                        & 4                        & 0.2                           & \multicolumn{2}{c|}{timeout}                                 & 12                       & 0.2                           & 953.7                       \\ \cline{3-3} \cline{6-14} 
		&                                                                                    & Overall      &                                             &                                           & 4   & 0.8                      & 6                        & 0.6                           & \multicolumn{2}{c|}{timeout}                                 & 7                        & 0.6                   & 282.8                       \\ \cline{2-3} \cline{6-14} 
		& \multirow{3}{*}{\begin{tabular}[c]{@{}l@{}}CdPlayer\\ ModeManager\end{tabular}}    & ModeManager  &                                             &                                           & 1   & \textbf{1}                        & 4                        & \textbf{1}                    & 10.9                        & 32.2                           & 4                        & \textbf{1}                           & 416.1                       \\ \cline{3-3} \cline{6-14} 
		&                                                                                    & InOn         &                                             &                                           & 1   & \textbf{1}                         & 5                        & 0.7                           & \multicolumn{2}{c|}{timeout}                                 & 5                        & 0.8                           & 876.9                       \\ \cline{3-3} \cline{6-14} 
		&                                                                                    & Overall      &                                             &                                           & 1   & \textbf{1}               & 2                        & \textbf{1}                    & 4.7                         & 17.5                           & 2                        & \textbf{1}                    & 138                         \\ \hline
		\multirow{3}{*}{ModelingALaunchAbortSystem}                                                                            & \multirow{2}{*}{Abort}                                                             & InabortLogic & \multirow{3}{*}{6}                          & \multirow{3}{*}{22}                       & 2   & \textbf{1}                       & 6                        & \textbf{1}                    & 518.9                       & 3.5                            & \multicolumn{3}{c|}{seg fault}                                                         \\ \cline{3-3} \cline{6-14} 
		&                                                                                    & Overall      &                                             &                                           & 1   & \textbf{1}                      & 4                        & \textbf{1}                    & 7.6                         & 39.1                           & 4                        & \textbf{1}                    & 70.6                        \\ \cline{2-3} \cline{6-14} 
		& \multicolumn{2}{l|}{ModeLogic}                                                                    &                                             &                                           & 4   & \textbf{1}               & 5                        & \textbf{1}                    & 52.2                        & 30.8                           & 5                        & 0.4                           & 107.8                       \\ \hline
		\multirow{2}{*}{\begin{tabular}[c]{@{}l@{}}ModelingAnIntersectionOf\\ Two1wayStreetsUsingStateflow\end{tabular}}       & \multicolumn{2}{l|}{InRed}                                                                        & \multirow{2}{*}{11}                         & \multirow{2}{*}{60}                       & 1   & 0.8                      & 8                        & 0.4                           & \multicolumn{2}{c|}{timeout}                                 & 8                        & 0.4                           & 105.6                       \\ \cline{2-3} \cline{6-14} 
		& \multicolumn{2}{l|}{Overall}                                                                      &                                             &                                           & 1   & \textbf{1}                      & 6                        & 0.6                           & \multicolumn{2}{c|}{timeout}                                 & 6                        & 0.6                    & 81.8                        \\ \hline
		\multicolumn{3}{|l|}{ModelingARedundantSensorPairUsingAtomicSubchart}                                                                                                                                                      & 6                                           & 20                                        & 4   & \textbf{1}                      & 4                        & \textbf{1}                    & 1007.6                      & 1.8                            & 5                        & 0.5                           & 72.4                        \\ \hline
		\multirow{6}{*}{ModelingASecuritySystem}                                                                               & \multirow{2}{*}{InAlarm}                                                           & InOn         & \multirow{6}{*}{16}                         & \multirow{6}{*}{100}                      & 16  & \textbf{1}                        & 4                        & \textbf{1}                    & 1599                        & 18.5                           & \multicolumn{3}{c|}{seg fault}                                                         \\ \cline{3-3} \cline{6-14} 
		&                                                                                    & Overall      &                                             &                                           & 1   & \textbf{1}               & 3                        & \textbf{1}                    & 7.4                         & 20.7                           & \multicolumn{3}{c|}{seg fault}                                                         \\ \cline{2-3} \cline{6-14} 
		& \multicolumn{2}{l|}{InDoor}                                                                       &                                             &                                           & 1   & \textbf{1}               & 3                        & \textbf{1}                    & 7.1                         & 16.7                           & \multicolumn{3}{c|}{seg fault}                                                         \\ \cline{2-3} \cline{6-14} 
		& \multirow{2}{*}{InMotion}                                                          & InActive     &                                             &                                           & 9   & \textbf{1}                        & 4                        & \textbf{1}                    & 1017.1                      & 8.9                            & \multicolumn{3}{c|}{seg fault}                                                         \\ \cline{3-3} \cline{6-14} 
		&                                                                                    & Overall      &                                             &                                           & 1   & \textbf{1}               & 3                        & \textbf{1}                    & 7.5                         & 15                             & \multicolumn{3}{c|}{seg fault}                                                         \\ \cline{2-3} \cline{6-14} 
		& \multicolumn{2}{l|}{InWin}                                                                        &                                             &                                           & 1   & \textbf{1}               & 3                        & \textbf{1}                    & 8.1                         & 17.8                           & \multicolumn{3}{c|}{seg fault}                                                         \\ \hline
		\multicolumn{3}{|l|}{MonitorTestPointsInStateflowChart}                                                                                                                                                                    & 2                                           & 20                                        & 1   & \textbf{1}               & 2                        & \textbf{1}                    & 2.9                         & 30.2                           & 2                        & \textbf{1}                    & 29.6                        \\ \hline
		\multicolumn{3}{|l|}{MooreTrafficLight}                                                                                                                                                                                    & 3                                           & 40                                        & 3   & \textbf{1}               & 7                        & \textbf{1}                    & 89.3                        & 38.4                           & 9                        & 0.7                    & 124                         \\ \hline
		\multicolumn{3}{|l|}{ReuseStatesByUsingAtomicSubcharts}                                                                                                                                                                    & 2                                           & 10                                        & 1   & \textbf{1}               & 3                        & \textbf{1}                    & 5.8                         & 27.3                           & 3                        & \textbf{1}                    & 52.8                        \\ \hline
		\multicolumn{3}{|l|}{SchedulingSimulinkAlgorithmsUsingStateflow}                                                                                                                                                           & 3                                           & 127                                       & 5   & \textbf{1}               & 3                        & \textbf{1}                    & 54.7                        & 17.7                           & 4                        & 0.8                    & 35.9                        \\ \hline
		\multicolumn{3}{|l|}{SequenceRecognitionUsingMealyAndMooreChart}                                                                                                                                                           & 2                                           & 30                                        & 1   & \textbf{1}               & 5                        & \textbf{1}                    & 9.8                         & 54.4                           & 5                        & \textbf{1}                    & 88.2                        \\ \hline
		\multicolumn{3}{|l|}{ServerQueueingSystem}                                                                                                                                                                                 & 4                                           & 40                                        & 2   & \textbf{1}               & 3                        & \textbf{1}                    & 13.8                        & 31                             & 4                        & 0.6                           & 99.3                        \\ \hline
		\multicolumn{3}{|l|}{StatesWhenEnabling}                                                                                                                                                                                   & 2                                           & 30                                        & 1   & \textbf{1}               & 4                        & \textbf{1}                    & 4.3                         & 35.8                           & 4                        & \textbf{1}                    & 32.1                        \\ \hline
		\multicolumn{3}{|l|}{StateTransitionMatrixViewForStateTransitionTable}                                                                                                                                                     & 3                                           & 25                                        & 4   & \textbf{1}               & 5                        & \textbf{1}                    & 53.9                        & 52.8                           & 7                        & 0.8                    & 89.1                        \\ \hline
		\multirow{2}{*}{Superstep}                                                                                             & \multicolumn{2}{l|}{With Super Step}                                                              & \multirow{2}{*}{1}                                           & \multirow{2}{*}{10}                                        & 1   & \textbf{1}               & 1                        & \textbf{1}                    & 139.7                       & 0.4                            & 1                        & \textbf{1}                    & 21.8                        \\ \cline{2-3} \cline{6-14} 
		& \multicolumn{2}{l|}{Without Super Step}                                                           &                                          &                                         & 1   & \textbf{1}               & 3                        & \textbf{1}                    & 141.4                       & 0.8                            & 3                        & \textbf{1}                     & 25.5                        \\ \hline
		\multicolumn{3}{|l|}{TemporalLogicScheduler}                                                                                                                                                                               & 2                                           & 202                                       & 6   & \textbf{1}               & 4                        & \textbf{1}                    & 270.8                       & 4.4                            & 4                        & \textbf{1}                    & 36                          \\ \hline
		\multicolumn{3}{|l|}{UsingSimulinkFunctionsToDesignSwitchingControllers}                                                                                                                                                   & 3                                           & 10                                        & 1   & \textbf{1}               & 4                        & \textbf{1}                    & 7.2                         & 27.1                           & 4                        & \textbf{1}                    & 41.5                        \\ \hline
		\multirow{2}{*}{VarSize}                                                                                               & \multicolumn{2}{l|}{SizeBasedProcessing}                                                          & \multirow{2}{*}{4}                          & \multirow{2}{*}{35}                       & 2   & \textbf{1}               & 3                        & \textbf{1}                    & 115.6                       & 3.4                            & 4                        & 0.6                    & 38.9                        \\ \cline{2-3} \cline{6-14} 
		& \multicolumn{2}{l|}{VarSizeSignalSource}                                                          &                                             &                                           & 2   & \textbf{1}               & 5                        & \textbf{1}                    & 157                         & 5.3                            & 6                        & 0.7                    & 47.4                        \\ \hline
		\multicolumn{3}{|l|}{ViewDifferencesBetweenMessagesEventsAndData}                                                                                                                                                          & 2                                           & 10                                        & 2   & \textbf{1}               & 4                        & \textbf{1}                    & 9                           & 46.9                           & 4                        & \textbf{1}                    & 34.8                        \\ \hline
		\multirow{3}{*}{YoYoControlOfSatellite}                                                                                & \multirow{2}{*}{InActive}                                                          & InReelMoving & \multirow{3}{*}{8}                          & \multirow{3}{*}{10}                       & 2   & \textbf{1}                      & 4                        & \textbf{1}                    & 18.5                        & 47.3                           & 4                        & \textbf{1}                    & 60.1                        \\ \cline{3-3} \cline{6-14} 
		&                                                                                    & Overall      &                                             &                                           & 2   & \textbf{1}                        & 4                        & \textbf{1}                    & 19.5                        & 46.7                           & 4                        & \textbf{1}                    & 71.9                        \\ \cline{2-3} \cline{6-14} 
		& \multicolumn{2}{l|}{Overall}                                                                      &                                             &                                           & 1   & \textbf{1}               & 3                        & \textbf{1}                    & 3.7                         & 28.7                           & 3                        & \textbf{1}                    & 43                          \\ \hline
	\end{tabular}
\end{table*}

\subsection{Experiments and Results}

For each benchmark, we generate an initial set of $50$ traces, each of length
$50$, by executing the system with randomly sampled inputs. We assume that the
$k$ value for counterexample validity check is known a priori and is supplied to
the algorithm for each benchmark. Some of the Stateflow models are implemented
as multiple parallel and hierarchical FSAs. For a given implementation
$\mathcal{S}$ and a set of observables $X$, we attempt to reproduce each state
machine separately using traces defined over all variables in $X$. We therefore generate
an abstraction with state transitions at a system level for each FSA in
$\mathcal{S}$.

The results are summarised in Table~\ref{tab:result}. We quantitatively assess
the quality of the final generated model for each FSA by assigning a score $d$,
computed as the fraction of state transitions in the Stateflow model that match
corresponding transitions in the abstraction. For hierarchical Stateflow models,
we flatten the FSAs and compare the learned abstraction with the flattened FSA.
We record the number of model learning iterations $i$, the number of states $N$
and degree of completeness $\alpha$ for the final model, the total runtime $T$
and the percentage of total runtime attributed to model learning, denoted by
$\%T_m$. We set a timeout of $10$\,h for our experiments. For benchmarks that
time out, we present the results for the model generated right before timeout.

\subsubsection{Runtime}
The active learning algorithm is able to generate abstractions in under $1$\,h
for the majority of the benchmarks. For the benchmarks that time out, we see
that the model checker tends to go through a large number of invalid
counterexamples before arriving at a valid counterexample for a condition
violation. This is because, depending on the size of the domain $D$ for the
variables $x \in X$, there can be a large number of possible valuations that
violate an extracted condition, of which very few may correspond to a valid
system state. In such cases, runtime can be improved by strengthening the
assumption $r$ in Fig.~\ref{fig:verify} with domain knowledge to guide the model
checker towards valid counterexamples. For the \emph{FrameSyncController}
benchmark, CBMC takes a long time to check each condition, even with $k = 1$.
This is because the implementation features several operations, such as memory
access and array operations, that especially increase proof complexity and proof
runtime.

\subsubsection{Generated Model Accuracy}
The algorithm is guaranteed to generate an abstraction that admits all system
behaviours, as is confirmed by $\alpha$ in Table~\ref{tab:result}. We also see
that $d = 1$ for these benchmarks. For two benchmarks, although the Simulink
model matched the generated abstraction $(d = 1)$, the algorithm timed out
before it could eliminate all spurious violations $(\alpha < 1)$.

\subsubsection{Number of Learning Iterations} 
In each learning iteration $j$, $|L(\mathcal{M}_j)| > |L(\mathcal{M}_{j-1})|$ as
$L(\mathcal{M}_j) \supseteq L(\mathcal{M}_{j{-}1}) \cup
\mathcal{T}_{\mathit{CE}_j}$ and $\mathcal{T}_{\mathit{CE}_j} \cap
L(\mathcal{M}_{j-1}) = \emptyset$. Here, $\mathcal{M}_j$ and
$\mathcal{T}_{\mathit{CE}_j}$ are the generated abstraction and the set of new
traces collected in iteration $j$ respectively. The algorithm terminates when
$L(\mathcal{M}_j) \supseteq \mathit{Traces}_X(S)$. The number of learning
iterations therefore depends on $|\mathit{Traces}_X(S) \cap L(\mathcal{M}_0)|$,
where $\mathcal{M}_0$ is the abstraction generated from the initial trace set.

\footnotetext[2]{The dataset includes another implementation of this benchmark
	with similar results. Due to space constraints, we present the results for only
	one of them.}

\subsection{Comparison with Random Sampling}\label{sec:random}
We performed a set of experiments to check if random sampling is sufficient to
learn abstractions that admit all behaviours. A~million randomly sampled inputs
are used to execute each benchmark. Generated traces are fed to T2M to passively
learn a model. T2M fails to generate a model~for~$7$ benchmarks, as its
predicate synthesis procedure returns `segmentation fault'. For $50\%$ of the
remaining benchmarks, random sampling fails to produce a model admitting all
system~behaviours~($\alpha < 1$).

\subsection{Threats to Validity}
The key threat to the validity of our experimental claim is benchmark bias. We
have attempted to limit this bias by using a set of benchmarks that was curated
by others.  Further, we use C implementations of Simulink Stateflow models that
are auto-generated using a specific code generator.  Although there is diversity
among these benchmarks, our algorithm may not generalise to software that is not
generated from Simulink models, or software generated using a different
code~generator.

\section{Related Work}
State-merge~\cite{Biermann:1972:SFM:1638603.1638997} is a popular approach for
learning finite automata from system traces. The approach is predominantly
passive and generated abstractions admit only those system behaviours
exemplified by the traces~\cite{edsm, Heule_2010, model_daikon,
	compute_walkinshaw, Walkinshaw2016}.

One of the earliest active model learning algorithms using state-merge is
Query-Driven State Merging (QSM)~\cite{qsm}, where model refinement is guided by
responses to membership queries posed to an end-user. Other active versions of
state-merge use model checking~\cite{state_merge, exact_fsm} and model-based
testing~\cite{state_merge_testing} to identify spurious behaviours in the
generated model. However, the learned model is not guaranteed to admit all
system behaviour. 


Angluin's L* algorithm~\cite{lstar} is a classic active automata learning
approach to construct Deterministic Finite Automata (DFA) for regular languages.
The approach assumes the presence of a Minimally Adequate Teacher (MAT) that has
sufficient information of the system to answer membership and equivalence
queries posed by the learning framework. Algorithms based on this MAT
framework~\cite{lstar,mat_star,ttt,lstar_mealy, active_kearns} can, in
principle,  generate exact system models. But the absence of an equivalence
oracle, in practice, often restricts their ability to generate exact models or
even accurate system over-approximations.

In a black-box setting, membership queries are posed as \textit{tests} on the
(unknown) System Under Learning (SUL). The elicited response to a test is used
to classify the corresponding query as accepting or rejecting. Equivalence
queries are often approximated using techniques such as conformance testing or
random testing~\cite{sl_star,tlv,ralib,black_box_cegar,dynamic_test}, through a
finite number of membership queries. An essential pre-requisite to enable
black-box model learning is that the SUL can be simulated with an input sequence
to elicit a response or output. Moreover, obtaining an adequate approximation of
an equivalence oracle in a black-box setting may require a large number of
membership queries, that is exponential in the number of states in the SUL. The
resulting high query complexity constrains these algorithms to learning only
partial models for large systems~\cite{Howar2019,Howar2018ActiveAL}.

One way to address these challenges is to combine model learning with white-box
techniques, such as fuzzing~\cite{fuzzing}, symbolic
execution~\cite{component_interface,component_interface_imprv} and model
checking~\cite{lstar_assume, lstar_assume_prob}, to extract system information
at a lower cost~\cite{Howar2019}. In~\cite{fuzzing}, model learning is combined
with mutation based testing that is guided by code coverage. This proves to be
more effective than conformance testing, but the approach does not always
produce complete models.
In~\cite{component_interface,component_interface_imprv}, symbolic execution is
used to answer membership queries and generate component interface abstractions
modeling safe orderings of component method calls. Sequences of method calls in
a query are symbolically executed to check if they reach an a priori known
unsafe state. However, learned models may be partial as method call orderings
that are unsafe but unknown due to insufficient domain knowledge are missed by
the approach. In~\cite{lstar_assume, lstar_assume_prob}, model checking is used
in combination with model learning for assume guarantee reasoning. The primary
goal of the approach is not to generate an abstract model of a component and may
therefore terminate before generating a complete model. Also, learned models are
defined over an a priori known finite alphabet consisting of observable actions.

Very closely related to our work are the algorithms that use L* in combination
with black-box testing~\cite{lstar_Peled1999} and model
checking~\cite{lstar_model, lstar_model2}. The latter uses pre-defined LTL
system properties, similar to~\cite{state_merge, exact_fsm}, and therefore
generated abstractions may not model system behaviours outside the scope of
these LTL properties. Black-box testing can be adopted to check degree of
completeness by simulating the learned model with a set of system execution
traces to identify missing behaviour. However, generated models are not
guaranteed to admit all system behaviour, as this requires a system load that
exercises the implementation to cover all behaviours.

An open challenge with query-based active model learning is learning symbolic
models. Many practical applications of L*~\cite{kroenig-lstar,lstar_assume} and
its variants are limited to learning system models defined over a finite
alphabet consisting of Boolean events, such as function calls, that need to be
known a priori. Sigma*~\cite{sigma*} addresses this by extending the L*
algorithm to learn symbolic models of software. Dynamic symbolic execution is
used to find constraints on inputs and expressions generating output to build a
symbolic alphabet. However, behaviours modeled by the generated abstraction are
limited to input-output steps of a software.

Another extension of the L* algorithm~\cite{mat_star} generates symbolic models
using membership and equivalence oracles. Designing and implementing such
oracles to answer queries on long system traces comprising sequences of
valuations of multiple variables, some of which could have large domains, is not
straightforward~\cite{Howar2019}. In~\cite{mapper}, manually constructed mappers
abstract concrete valuations into a finite symbolic alphabet. However, this
process can be laborious and error prone. In~\cite{abstract_alphabet}, this
problem is overcome using automated alphabet abstraction refinement.
In~\cite{symbolic_mealy},  an inferred mealy machine is converted to a symbolic
version in a post-processing step. These algorithms are however restricted to
learning models with simple predicates such as equality/inequality relations.

The SL* algorithm~\cite{sl_star} extends MAT model learning to infer register
automata that model both control flow and data flow. In addition to
equality/inequality relations, automaton transition feature simple arithmetic
expressions such as increment by $1$ and sum. Due to the high query complexity
it is not obvious how the approach can be generalised to symbolic models over
richer theories. An extension of this algorithm~\cite{grey_box_sl}, uses taint
analysis to boost performance by extracting constraints on input and output
parameters. Both algorithms use individual oracles for each type of operation
represented in the symbolic model and do not allow analysis of multiple or more
involved operations on data values. The SUL is modeled as a register automaton
and model learning is performed in a black-box setting, thereby generating
partial models.

In our approach, the procedure used to check degree of completeness for the
learned model operates at the level of the abstraction and not system traces,
and therefore can be easily implemented using existing model checkers. Further,
the model learning procedure and subsequent evaluation of the degree of
completeness are independent of each other. This enables our approach to generate
more expressive models, when combined with a model learning component that can
infer transition guards and data update functions from traces, such as
T2M~\cite{jeppu}.
\section{Use-Cases and Future Work} 

In this paper, we have presented a new active model-learning algorithm to
learning FSA abstractions of a system implementation from traces. The generated
models are guaranteed to admit all system behaviour over a set of observable
variables.

This can be particularly useful when system specifications are incomplete, and
so any implementation errors outside the scope of defined requirements cannot be
flagged. This is a common risk when essential domain knowledge gets
progressively pruned as it is passed on from one team to another during the
development life cycle. In such scenarios, manual inspection of~the learned
models can help identify errors in implementation. With our approach, the
conditions extracted from the learned model are  invariants that hold on the
implementation. These can be used as additional specifications to verify
multiple system implementations. The approach can also be used to evaluate~test
coverage for a given test suite and generate new tests to address coverage
holes.

In the future, we intend to explore these potential use-cases further. This will
drive improvements to reduce runtime, such as ways to guide the model checker
towards valid counterexamples. We intend also to investigate extensions of the
approach to model recursive state machines.

\bibliographystyle{IEEEtran}
\bibliography{IEEEabrv,paper}

\end{document}